\documentclass{osa-article}
\journal{oe}

\usepackage{lineno}
\usepackage{caption}
\usepackage{mathtools}
\usepackage{amsmath}
\usepackage[subrefformat=parens]{subcaption}
\usepackage{siunitx}
\usepackage{physics}

\usepackage{soul,xcolor}

\newcommand{\add}[1]{\textcolor{black}{#1}}

\graphicspath{{./pictures/}}

\begin{document}

\title{Programmable time-multiplexed squeezed light source}

\author{Hiroko Tomoda,\authormark{1} Takato Yoshida,\authormark{1} Takahiro Kashiwazaki,\authormark{2} Takeshi Umeki,\authormark{2} Yutaro Enomoto,\authormark{1} and Shuntaro Takeda\authormark{1,*}}

\address{\authormark{1}Department of Applied Physics, School of Engineering, The University of Tokyo, 7-3-1 Hongo, Bunkyo-ku, Tokyo 113-8656, Japan\\
\authormark{2}NTT Device Technology Labs, NTT Corporation, 3-1, Morinosato Wakamiya, Atsugi, Kanagawa 243-0198, Japan}

\email{\authormark{*}takeda@ap.t.u-tokyo.ac.jp} %% email address is required



\begin{abstract}
	One of the leading approaches to large-scale quantum information processing (QIP) is the continuous-variable (CV) scheme based on time multiplexing (TM).
	As a fundamental building block for this approach, quantum light sources to sequentially produce time-multiplexed squeezed-light pulses are required; however, conventional CV TM experiments have used fixed light sources that can only output the squeezed pulses with the same squeezing levels and phases.
	We here demonstrate a programmable time-multiplexed squeezed light source that can generate sequential squeezed pulses with various squeezing levels and phases at a time interval below \SI{100}{\nano\second}.
	The generation pattern can be arbitrarily chosen by software without changing its hardware configuration.
	This is enabled by using a waveguide optical parametric amplifier and modulating its continuous pump light.
	Our light source will implement various large-scale CV QIP tasks.
\end{abstract}

%%%%%%%%%%%%%%%%%%%%%%%%%%  body  %%%%%%%%%%%%%%%%%%%%%%%%%%
\section{Introduction}
Nowadays, quantum information processing (QIP) attracts much attention and is implemented with various physical systems.
Among them, optical systems are one of the leading candidates since they have a long coherence time even at room temperature and atmospheric pressure.
Optical QIP has various applications such as quantum computing, quantum sensing, and quantum communication~\cite{flamini_photonic_2019,pirandola_advances_2018}.
Moreover, optical QIP has recently been scaled up dramatically by the promising approach of combining the continuous-variable (CV) scheme and the time-multiplexing (TM) scheme~\cite{takeda_toward_2019}.
The CV scheme has the advantage that we can deterministically synthesize entanglement and perform quantum operations using deterministically prepared squeezed states.
On the other hand, the TM scheme can easily handle a large amount of quantum information by encoding it in a series of optical pulses arranged in the time domain.
By combining these two schemes, we can realize large-scale QIP with an unlimited number of squeezed pulses propagating in a few optical paths, in principle.
Such high scalability of the CV TM scheme has already been demonstrated in several experiments, such as the generation of large-scale quantum entanglement~\cite{yokoyama_ultra-large-scale_2013,yoshikawa_invited_2016,larsen_deterministic_2019,asavanant_generation_2019}, the multi-step quantum computation based on the one-way~\cite{larsen_deterministic_2021,asavanant_time-domain-multiplexed_2021} or loop-based models~\cite{takeda_universal_2017,takeda_-demand_2019,enomoto_programmable_2021}, and even the demonstration of quantum supremacy by Gaussian boson sampling (GBS)~\cite{madsen_quantum_2022}.

In the CV TM scheme, quantum light sources to sequentially produce time-multiplexed quantum-state pulses are required to prepare initial states for QIP tasks.
In particular, the time-multiplexed squeezed light sources are the key enabler for various QIP tasks since squeezed states are the basis of entanglement synthesis and quantum operations.
In general, each pulse should have a different squeezed state when we perform various QIP tasks.
However, the previous CV TM experiments have used optical parametric oscillators (OPOs) with constant pump power to produce squeezed pulses with the same squeezing levels and phases~\cite{yokoyama_ultra-large-scale_2013,yoshikawa_invited_2016,larsen_deterministic_2019,asavanant_generation_2019,larsen_fiber-coupled_2019,larsen_deterministic_2021,asavanant_time-domain-multiplexed_2021,takeda_-demand_2019,enomoto_programmable_2021,madsen_quantum_2022}.
Such fixed squeezed light sources severely limit the possible QIP tasks.
For example, this limitation is a critical issue in loop-based quantum computing~\cite{takeda_universal_2017}, GBS~\cite{hamilton_gaussian_2017}, and quantum reservoir computing~\cite{garcia-beni_scalable_2022}.

\begin{figure}[htp]
	\centering
	\includegraphics[width=7cm]{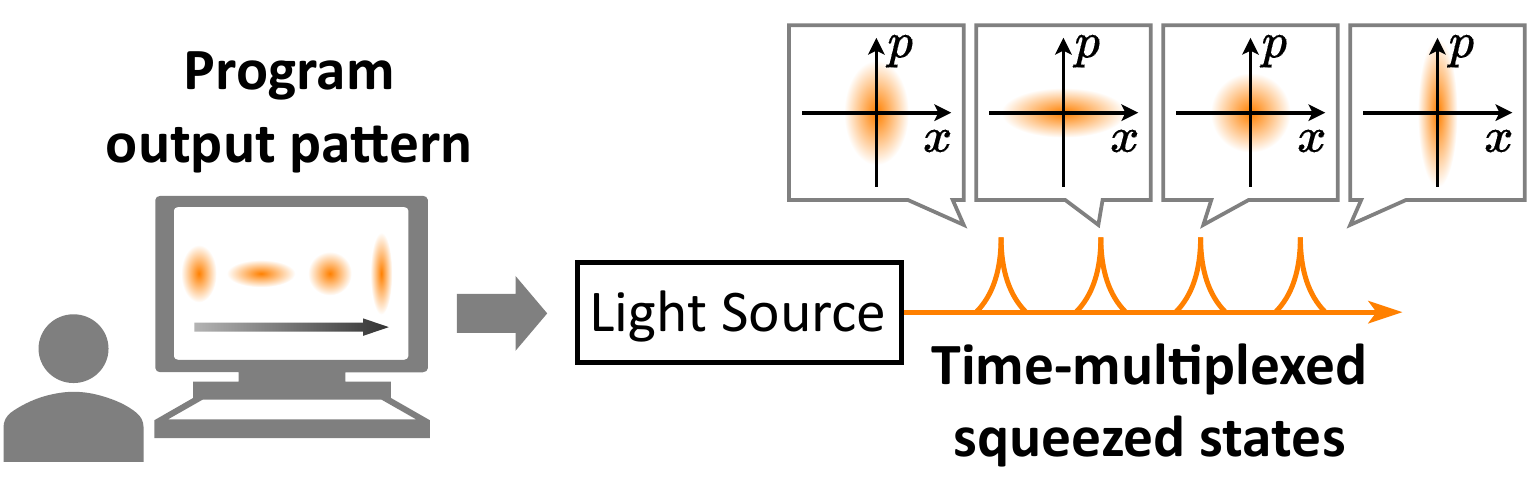}
	\caption{Schematic image of programmable time-multiplexed squeezed light source.
	This light source can produce sequential squeezed pulses with the different squeezing levels and phases by a computer program without the experimental setup change.}
	\label{pic:schematic}
\end{figure}

This limitation can be overcome by a programmable time-multiplexed squeezed light source in Fig.~\ref{pic:schematic} that meets the following two requirements.
The first requirement is the ability to sequentially produce time-multiplexed squeezed pulses by changing the squeezing levels and phases pulse by pulse.
The pulse interval should be adjustable among several tens to hundreds of nanoseconds, which are the typical time scales of previous CV TM experiments~\cite{yokoyama_ultra-large-scale_2013, yoshikawa_invited_2016,larsen_deterministic_2019, asavanant_generation_2019, larsen_fiber-coupled_2019, larsen_deterministic_2021,asavanant_time-domain-multiplexed_2021,takeda_-demand_2019,enomoto_programmable_2021,madsen_quantum_2022}.
The second requirement is the ability to program the pattern of such a train of squeezed pulses by software so that we can perform various QIP tasks without changing the hardware configuration.
One way to construct such a light source is to operate multiple OPOs in parallel and time-multiplex these output states into a single optical path by an optical switch.
However, this configuration requires numerous resources and additional switching loss that degrades the loss-sensitive squeezed states.

Here, we realize a programmable time-multiplexed squeezed light source by using a waveguide optical parametric amplifier (OPA), instead of an OPO, and pumping it with a temporally modulated continuous pump light.
%\erase{In contrast to the cavity-based OPO, the waveguide OPA has a single-path structure.}
%\erase{Thus, the variation of the pump power and phase instantaneously affects the squeezing level and phase of the output squeezed light.}
\add{In contrast to the cavity-based OPO, the waveguide OPA has long been studied as an alternative means to produce squeezing without cavity ~\cite{anderson_quadrature_1995, serkland_squeezing_1995, yoshino_generation_2007, eto_observation_2007, kaiser_fully_2016, nehra_few-cycle_2022}.}
\add{Due to its single-path structure, the variation of the pump power and phase instantaneously affects the squeezing level and phase of the output squeezed light.}
As a result, we can rapidly change the squeezing level and phase pulse by pulse through the modulation of the pump light.
This pump modulation is determined by an arbitrary waveform generator (AWG) signal.
Therefore, the output pattern of the squeezed pulses is easily programmable.
This configuration is compact and does not introduce any additional loss to the squeezed states since the output switching is realized through the modulation of the classical pump light.
\add{Thus far, squeezed light sources multiplexed in frequency or spatial modes have been also developed~\cite{pan_orbital-angular-momentum_2019,liu_orbital_2020,li_deterministic_2020,chen_experimental_2014,roslund_wavelength-multiplexed_2014} and used for some QIP tasks, but our time-multiplexed source will enable more variety of QIP tasks in a programmable and scalable way.}
In this paper, we experimentally demonstrate the programmable generation of various patterns of squeezed pulses at the time interval below \SI{100}{\nano\second}, which is comparable to the time scale of the previous CV TM experiments~\cite{yokoyama_ultra-large-scale_2013, yoshikawa_invited_2016,larsen_deterministic_2019, asavanant_generation_2019, larsen_fiber-coupled_2019, larsen_deterministic_2021,asavanant_time-domain-multiplexed_2021,takeda_-demand_2019,enomoto_programmable_2021,madsen_quantum_2022}.
This result shows the applicability of our light source to various CV TM experiments.
Furthermore, we demonstrate that our light source can even produce quantum entanglement in two frequency bins, revealing that this light source is also potentially useful for the frequency-multiplexing approach~\cite{pfister_continuous-variable_2020,roslund_wavelength-multiplexed_2014}.
Since we introduce a recently-developed waveguide OPA module which achieved record-high 6.3-\si{\deci\bel} squeezed light over the 6-\si{\tera\hertz} sideband in the previous study~\cite{kashiwazaki_fabrication_2021}, our light source is broadband and can potentially produce highly squeezed light, thus enabling the high-speed and high-fidelity QIP.
%\erase{Our light source can be applied to various QIP tasks including quantum computation~\protect{\cite{takeda_universal_2017,hamilton_gaussian_2017,garcia-beni_scalable_2022}}, quantum sensing~\cite{lawrie_quantum_2019}, and quantum communication~\cite{flamini_photonic_2019} in both the time and frequency domains, opening a new door to large-scale and programmable optical QIP.}
\add{Thus, our light source can potentially be applied to a wide variety of QIP tasks including quantum computing~\cite{takeda_universal_2017,hamilton_gaussian_2017,garcia-beni_scalable_2022}, quantum sensing~\cite{lawrie_quantum_2019}, and quantum communication~\cite{flamini_photonic_2019} in both the time and frequency domains, opening a new door to large-scale and programmable optical QIP.}
\add{In fact, although the squeezing levels demonstrated in this paper are not yet sufficient for QIP tasks based on high-fidelity quantum gates and highly entangled states, programmable generation of low squeezing in our system already has important applications, such as quantum computations and quantum state generation based on GBS~\cite{huh_boson_2015,sempere-llagostera_experimentally_2022,sabapathy_production_2019,fukui_efficient_2022}.}

This paper is organized as follows.
In Sec.~\ref{sec:Experimental setup}, we describe the details of our experimental setup.
In Sec.~\ref{sec:Results}, we show the experimental results.
In Sec.~\ref{sec:Squeezing spectrum of OPA}, we analyze the squeezing and anti-squeezing spectra of the output squeezed light of the waveguide OPA.
In Sec.~\ref{sec:Arbitrary waveform shaping and time variation of squeezing level}, we evaluate the pump light modulation system.
We observed the time variation of the pump light and quadrature variance of the OPA output by modulating the pump light to various waveforms.
In Sec.~\ref{sec:Generation of time-multiplexed squeezed states}, we demonstrate that our system can output squeezed pulses with different squeezing levels and phases programmably using the above modulation system.
We switched each squeezing level and phase of sequential wave-packet modes and evaluated them.
In Sec.~\ref{sec:Generation of frequency-entangled states}, to show that our light source can also output useful states in the frequency domain, we generate two-mode entangled states in the frequency domain.
In Sec.~\ref{sec:Coclusion}, we summarize this study and explain the prospects.

\section{Experimental setup}\label{sec:Experimental setup}
\begin{figure}[bp]
	\centering
	\includegraphics[width=13.2cm]{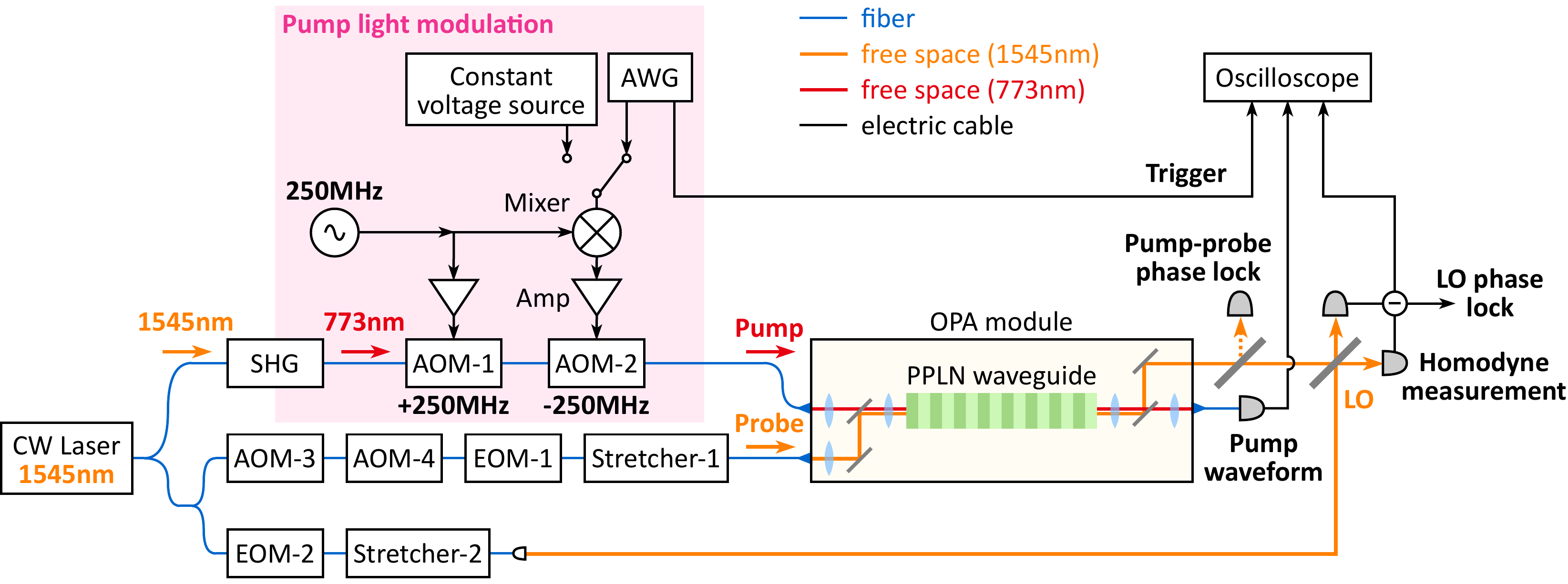}
	\caption{Experimental setup. 
	CW, continuous wave;
	SHG, second harmonic generator;
	AOM, acousto-optic modulator;
	EOM, electro-optic modulator;
	AWG, arbitrary waveform generator;
	OPA, optical parametric amplifier;
	PPLN, periodically poled lithium niobate;
	LO, local oscillator.
	}
	\label{pic:setup}
\end{figure}

Figure \ref{pic:setup} shows the experimental setup.
This system is divided into three parts: shaping the pump light to an arbitrary waveform, generating squeezed states from the modulated pump light, and evaluating the produced states.

In the first part, we use a continuous-wave (CW) laser at \SI{1545}{\nano\metre} and generate the second harmonic light at \SI{773}{\nano\metre}.
This light is then modulated by acousto-optic modulators (AOMs) before the OPA.
We avoid the frequency change of the light by using two AOMs: AOM-1 (AA OPTO-ELECTRONIC, MT250-NIR6-Fio-PM0,5-J1-A-VSF) for \num{250}-\si{\mega\hertz} upshift and AOM-2 (AA OPTO-ELECTRONIC, MT250-NIR6-Fio-PM0,5-J1-A-(s-)-VSF) for \num{250}-\si{\mega\hertz} downshift.
These AOMs have a rise time of \SI{6}{\nano\second} that is related to the time for acoustic waves to travel through an optical beam and this value characterizes an analog modulation bandwidth (BW, \SI{-3}{\deci\bel}) of \SI{80}{\mega\hertz}.
They are sufficiently fast to switch various squeezed states at the time interval of several tens to hundreds of nanoseconds.
Both AOMs are driven with a \SI{250}{\mega\hertz}-radio-frequency (RF) signal, and only AOM-2 driving signal is modulated.
Generally, by modulating the amplitude and phase of the AOM driving signal, we can arbitrarily modulate the amplitude and phase of the AOM output light.
In the present setup, we can control AOM-2 driving signal by an AWG (Keysight 81160A, BW \SI{330}{\mega\hertz}) with the assistance of a mixer.
The amplitude of AOM-2 driving signal can be arbitrarily changed with the AWG voltage amplitude, and the phase of that can be shifted by \SI{180}{\degree} with the AWG voltage sign.
If necessary, an arbitrary phase shift can be achieved by introducing an optical or electric phase modulator.

In the second part, the modulated light at \SI{773}{\nano\metre} is injected into the OPA.
This OPA is a fiber-coupled module to generate squeezed light with a periodically poled LiNbO$_3$ waveguide inside.
The same type module generated 6.3-\si{\deci\bel} squeezed light over the 6-\si{\tera\hertz} sideband in the previous experiment~\cite{kashiwazaki_fabrication_2021}.
This squeezing level is the best record observed in fully fiber-closed systems and is enabled by a low-loss PPLN waveguide in Ref.~\cite{kashiwazaki_fabrication_2021}, which has the following three features.
First, this waveguide is doped with ZnO and bonded directly onto a LiTaO$_3$ wafer.
This structure suppresses the pump-induced phenomena such as the photorefractive effect and consequently makes the waveguide durable against high pump power.
Second, the cross-section shape of this waveguide is designed for the quasi-single-mode light, which inhibits the contamination by the higher-order-mode light.
Third, this waveguide is made by a mechanically sculpturing process instead of a dry etching process~\cite{kashiwazaki_continuous-wave_2020}.
Then the sidewalls of the waveguide get smoother and the propagation loss becomes lower.
Since this waveguide OPA is a single-path squeezer, we can instantaneously reflect the pump modulation to the output squeezing.
This point is different from the conventional cavity-based squeezer, OPO.
In principle, we can switch the squeezing level and phase at speeds of over THz if the modulation speed of the pump light is not limited.
Note that the squeezing phase shift $\theta$ can be introduced by the pump phase shift $2\theta$~\cite{wu_squeezed_1987}.
In our setup, we can shift the squeezing phase by \SI{90}{\degree} when we invert the AWG voltage sign and shift the pump phase by \SI{180}{\degree}.

In the third part, the squeezed light is sent to free space, and its quadrature amplitude is measured with the homodyne detector (BW $\sim$\SI{200}{\mega\hertz}).
In addition, the pump power variation is monitored by the photodetector (BW \SI{1}{\giga\hertz}) after the OPA.
These data are acquired by the oscilloscope (BW \SI{1}{\giga\hertz}) at the trigger signal that is generated from the AWG together with the arbitrary waveform signal for the pump modulation.

In Fig.~\ref{pic:setup}, some apparatus is required for optical phase control.
To generate and observe squeezing, we have to fix the phase relationship between the pump light, probe light, and local oscillator (LO) light.
We alternately repeat the phase control period and measurement period at \SI{2}{\kilo\hertz}.
During the control period, we set the pump power constant by connecting a constant voltage source to the mixer and input the probe light by turning on AOM-3 and AOM-4.
We modulate the probe phase with an electro-optic modulator (EOM-1) and detect a small fraction of the probe light after the OPA.
This signal is demodulated and fed back to Stretcher-1 to lock the phase relationship between the pump and probe light.
Besides, the relative phase between the probe and LO light is fixed with Stretcher-2 by monitoring the homodyne signal.
By default, the LO phase is locked at the phase where we can measure the squeezed quadrature at positive AWG voltage.
During the measurement period, we block the probe light by turning off AOM-3 and AOM-4, and shift the LO phase with EOM-2 from the locked phase to the phase where we measure the quadratures.

\section{Results}\label{sec:Results}
\subsection{\add{Evaluation} of waveguide OPA}\label{sec:Squeezing spectrum of OPA}
\begin{figure}[htp]
	\centering
	\includegraphics{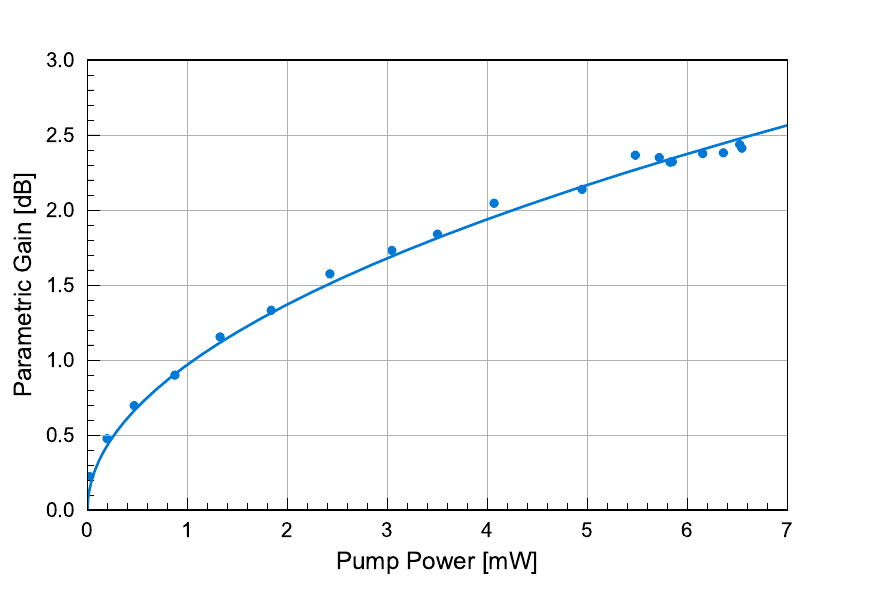}
	\caption{\add{Dependence of OPA's parametric gain on pump power.
	The pump power at the horizontal axis was measured at the output port of the OPA.}
	}
	\label{pic:parametric_gain}
% \end{figure}

% \begin{figure}[btp]
	\centering
	\includegraphics{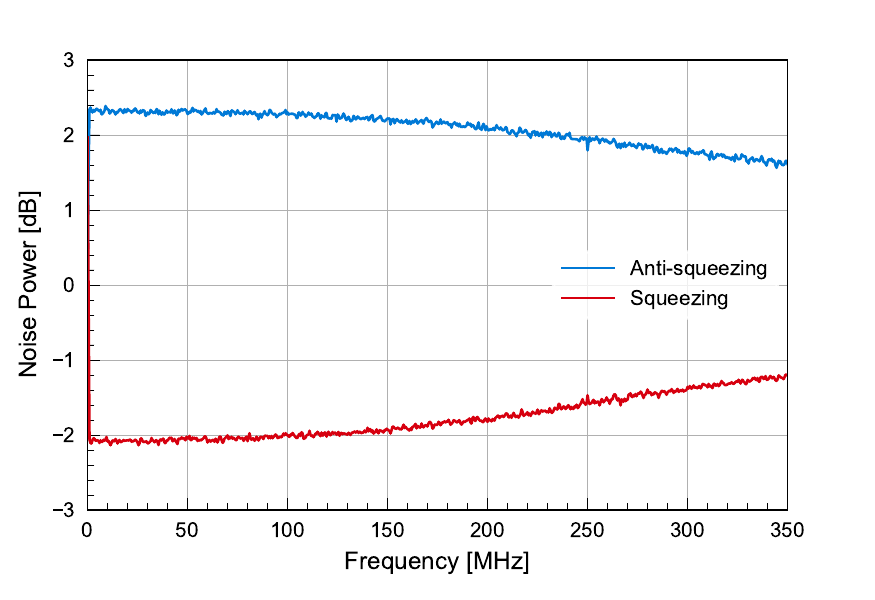}
	\caption{Squeezing and anti-squeezing spectra of output field of waveguide OPA.
	Noise power is normalized to the shot noise level.
	\add{This spectra was obtained from the time-series quadratures which were measured by the homodyne detector with \SI{200}{\mega\hertz} bandwidth.
	Due to this bandwidth, the observed squeezing and anti-squeezing levels decrease at higher frequencies although this type of OPA can generate squeezed light over \SI{6}{\tera\hertz}~\cite{kashiwazaki_fabrication_2021}.}
	}
	\label{pic:1-2-1}
\end{figure}

%\erase{To evaluate the performance of the waveguide OPA itself and the optical loss of our experimental setup, we first measured the squeezing spectrum of the OPA output field with the constant pump power of 15mW.}
%\erase{Here the pump power is limited to less than about 20mW due to the absolute maximum rating of input light power and maximum diffraction efficiency of AOM-1 and AOM-2 in our current setup.}
\add{To evaluate the performance of the waveguide OPA itself, we first measured the dependence of the OPA's parametric gain on the pump power.
This result is shown in Fig.~\ref{pic:parametric_gain}.
Note that the horizontal axis is the output pump power measured at the output port of the OPA (overall transmittance of the pump light through the OPA module is $40$\%).
In our setup, the output pump power is limited to \SI{\sim6.5}{\milli\watt} due to AOM-1 and AOM-2, which have absolute maximum ratings of input light power and maximum diffraction efficiencies.
In Fig.~\ref{pic:parametric_gain}, the fitting curve is obtained by the theoretical formulae~\cite{serkland_squeezing_1995}.}

\add{Next, we measured the squeezing spectrum of the OPA output field at the output pump power of \SI{6.5}{\milli\watt} and evaluated the optical loss of our experimental setup.}
We obtained \num{50000} frames of time-series waveforms of squeezed and anti-squeezed quadratures for each with the homodyne measurement.
We then calculated their spectra by a fast Fourier transform and plotted the average of them in Fig.~\ref{pic:1-2-1}.
This OPA module itself is expected to generate the squeezed light over \SI{6}{\tera\hertz}~\cite{kashiwazaki_fabrication_2021}, but the observed squeezing and anti-squeezing levels decrease at higher frequencies in Fig.~\ref{pic:1-2-1} due to the homodyne detector bandwidth, $\sim$\SI{200}{\mega\hertz}.

Using the low-frequency region of these spectra, we estimated the pure squeezing level and optical loss.
We split \num{50000} samples into 10 sets and analyzed each set in the following way to estimate the standard errors.
We first obtained the average values of the observed squeezing and anti-squeezing levels in the range of \SIrange{1}{10}{\mega\hertz}, respectively.
Based on these two values, we calculated the pure squeezing level before suffering from the loss as \SI[separate-uncertainty=true]{2.71 \pm 0.01}{\deci\bel}.
This result is consistent with the performance of the same type OPA~\cite{kashiwazaki_fabrication_2021} at the same pump power.
Simultaneously, the loss of our setup was calculated to be $18.3 \pm 0.5$\%.
It approximately corresponds to the total of each component loss: 9\% internal loss of the OPA module, 2\% propagation loss outside the OPA, 3\% mode-matching loss in homodyne measurement, and 1\% inefficiency of photodiodes.
Here the OPA internal loss 9\% was obtained from 5\% loss of the mirrors and lenses in the module and 4\% waveguide propagation loss, which was roughly estimated under the same assumption as Ref.~\cite{takanashi_4-db_2020}.
Note that the estimated loss value, 18.3\%, is used later when we calculate the theoretical quadrature variance.

\subsection{Arbitrary pump-waveform shaping and time variation of squeezing level}\label{sec:Arbitrary waveform shaping and time variation of squeezing level}

\begin{figure}[bp]
		\begin{minipage}[t]{0.49\linewidth}
			\centering
			\includegraphics{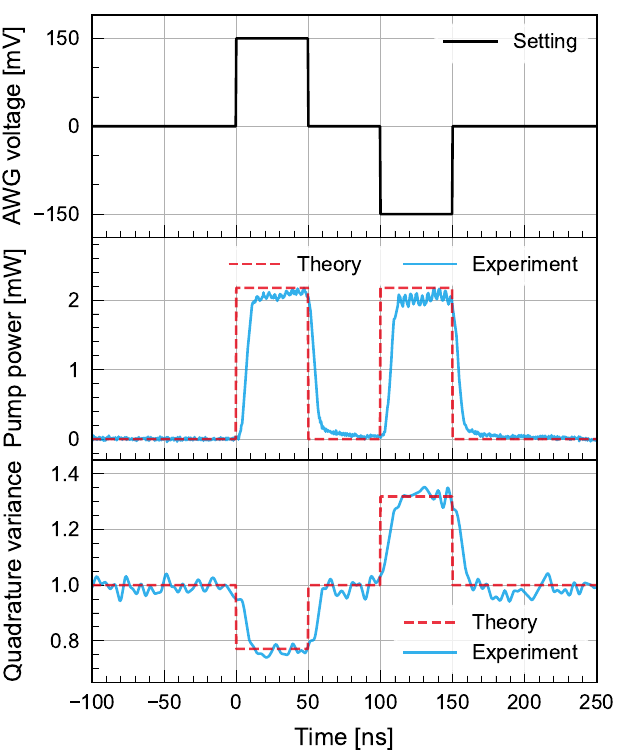}
			\subcaption{}\label{pic:2a}
		  \end{minipage}
		\begin{minipage}[t]{0.49\linewidth}
			\centering
			\includegraphics{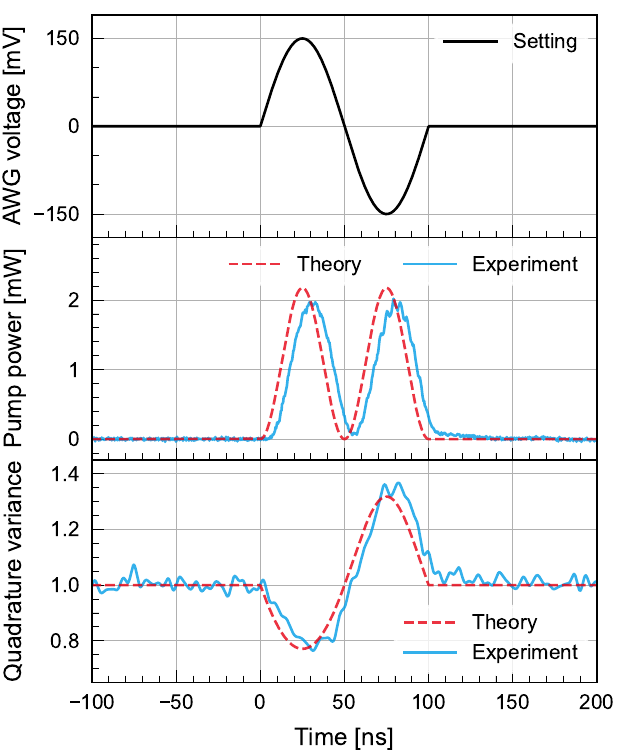}
			\subcaption{}\label{pic:2b}
		\end{minipage}
		\vspace{8pt}\\
		\begin{minipage}[t]{0.49\linewidth}
			\centering
			\includegraphics{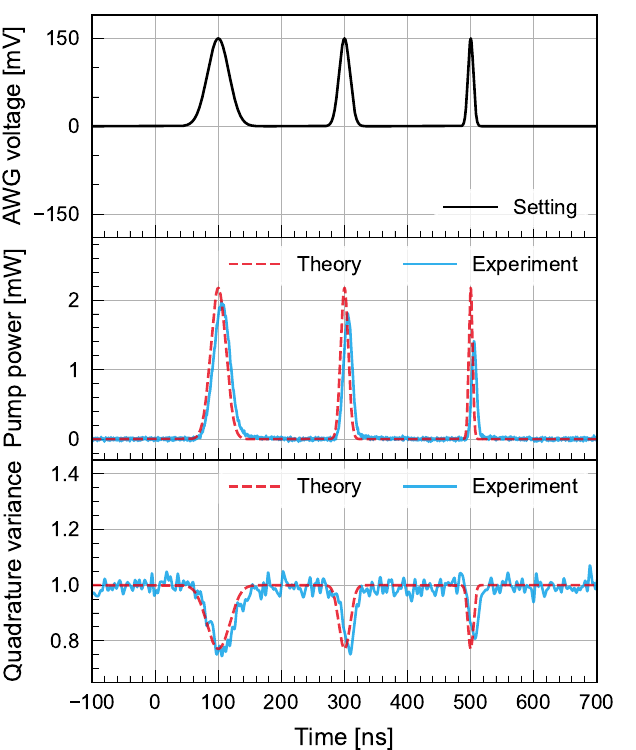}
			\subcaption{}\label{pic:2c}
		\end{minipage}
		\begin{minipage}[t]{0.49\linewidth}
			\centering
			\includegraphics{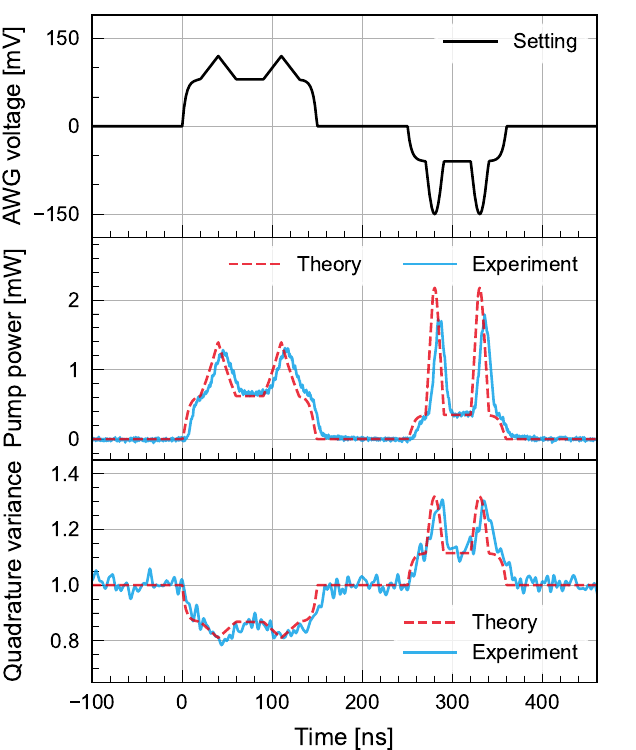}
			\subcaption{}\label{pic:2d}
		\end{minipage}  
	\caption{Arbitrary pump-waveform shaping and time variation of squeezing level.
	Upper panels of (a)-(d) show the voltage setting of the AWG.
	Middle panels show the time variation of the pump power, which was monitored after the OPA.
	Lower panels show the time variation of the quadrature variance, which was scaled by the vacuum variance.
	\add{Here the variance is calculated from the time-series quadratures which are filtered by a low-pass filter with \SI{100}{\mega\hertz} cutoff since our homodyne detector bandwidth is \SI{200}{\mega\hertz}.}
	We evaluate four different waveforms:
	(a) Square wave,
	(b) Sine wave (\SI{10}{\mega\hertz}),
	(c) Gaussian wave (FWHM is \SI{40}{\nano\second}, \SI{20}{\nano\second}, and \SI{10}{\nano\second} in order from the left), and
	(d) Cat and rabbit.
	}\label{pic:2}
\end{figure}

Next, we evaluated how fast and accurately we can modulate the pump light and thereby control the squeezing level and phase.
We measured the power of the pump light and the quadrature amplitude of the squeezed light by generating four different waveforms from the AWG as shown in Fig.~\ref{pic:2}.

In Fig.~\ref{pic:2}, the upper panels of (a)-(d) show the AWG output, and the middle panels show the time variation of the pump power.
The experimental pump power was converted from the output voltage of the photodetector monitoring the output pump light from the OPA.
The theoretical lines of the pump power were obtained as follows.
We assumed that the pump power is proportional to the RF power provided to AOM-2 and that this RF power is also proportional to the square of the AWG voltage.
These assumptions were experimentally confirmed as long as the absolute voltage of the AWG is below \SI{160}{\milli\volt}.
Thus, we fit the dependence of the pump power on the AWG voltage with a quadratic function of the AWG voltage.
We finally obtained the theoretical lines of pump power from the AWG setting voltage.
Now for the horizontal axis, we defined the start time of the AWG output as the axis origin.
We then adjusted the horizontal position of the experimental pump power so that their variation point matches that of the theoretical line in the middle panel of Fig.~\ref{pic:2a}, and also adjusted other middle panels with the same time shift.

Comparing the experimental results with the theoretical lines in the middle panels, we can evaluate our pump light modulation system.
From Fig.~\ref{pic:2a}, the rise time (10-90\%) of the pump power is about \SI{7}{\nano\second}, which is almost comparable to the above-mentioned AOM rise time, \SI{6}{\nano\second}.
In Fig.~\ref{pic:2b}, the pump power smoothly varies when the AWG signal smoothly varies.
Figure~\ref{pic:2c} shows how fast modulation our system can follow when we set three Gaussian pulses with different pulse widths.
Their full width at half maximum (FWHM) is \SI{40}{\nano\second}, \SI{20}{\nano\second}, and \SI{10}{\nano\second} in order from the left.
As the FWHM is smaller, the peaks of the pump pulses get lower due to the limited response speed of the modulation system.
Finally, Fig.~\ref{pic:2d} shows that we can generate the pump light with such a complex waveform signal.
Incidentally, the pump power sometimes oscillates at \SI{250}{\mega\hertz}, particularly in the upper flat region of the second square pulse in Fig.~\ref{pic:2a}.
It is attributed to the unwanted zeroth-order beam from AOM-1 and AOM-2, which is slightly coupled to the fiber and creates the beat signal with the first-order diffracted beam.

Lower panels in Fig.~\ref{pic:2} show the time variation of the quadrature variance.
We locked the LO phase at the phase where we can measure the squeezed quadrature at positive AWG voltage as mentioned in Sec.~\ref{sec:Experimental setup}.
In each of the four AWG settings, we obtained \num{5000} time-series waveforms of quadrature amplitudes.
We filtered these original time waveforms with a 255-tap finite-impulse-response low-pass filter at \num{100}-\si{\mega\hertz} cutoff, considering the homodyne detector bandwidth, $\sim$\SI{200}{\mega\hertz}.
We calculated pointwise quadrature variance for every data point from the filtered time-series data.
Finally, this variance was normalized by the vacuum variance acquired immediately before each measurement.
The theoretical values for the variances were also plotted based on the following procedure.
%\erase{In a separate preliminary measurement, we had measured the dependence of the OPA's parametric gain on the pump power.}
\add{As already mentioned in Sec.~\ref{sec:Squeezing spectrum of OPA}, we had measured the dependence of the OPA's parametric gain on the pump power (Fig.~\ref{pic:parametric_gain}).}
This result was well fitted by the curve of a theoretical model~\cite{serkland_squeezing_1995}.
Based on the fitting curve, we can estimate the parametric gain from the pump power.
Here we used the pump power that had been obtained from the AWG voltage as mentioned above.
As a result, we can estimate the parametric gain from the AWG voltage.
Using this parametric gain and loss value of 18.3\% in Sec.~\ref{sec:Squeezing spectrum of OPA}, we finally calculated the theoretical quadrature variance~\cite{serkland_squeezing_1995}.
As for the horizontal position, we adjusted the experimental data in the same way as the time variation of the pump power in the middle panels.

In Fig.~\ref{pic:2}, the quadrature variance is smaller than one at positive AWG voltage and larger at negative AWG voltage.
In our system, when we set positive AWG voltage, the squeezed quadrature is measured and thereby the quadrature variance is expected to be smaller than one.
On the other hand, at negative AWG voltage, the squeezing phase is shifted by \SI{90}{\degree} as mentioned in Sec.~\ref{sec:Experimental setup}.
Then the anti-squeezed quadrature is measured and the quadrature variance is expected to be larger than one.
We find that the experimental results are consistent with the theoretical ones and that their difference is attributed to the bandwidth of our pump light modulation system.

These results show that we can control the squeezing level and phase of the output squeezed light by modulating the AOM driving signal with the AWG and thereby modulating the pump light on $\sim$10-\si{\nano\second} time scale.
Since our modulation speed is limited by the AOM analog modulation bandwidth, it can be improved, for example, by using faster modulators such as EOMs~\cite{nie_active_2017}.

\subsection{Generation of time-multiplexed squeezed states}\label{sec:Generation of time-multiplexed squeezed states}

\begin{figure}[bp]
	\begin{minipage}[t]{\linewidth}
	  \centering
	  \includegraphics{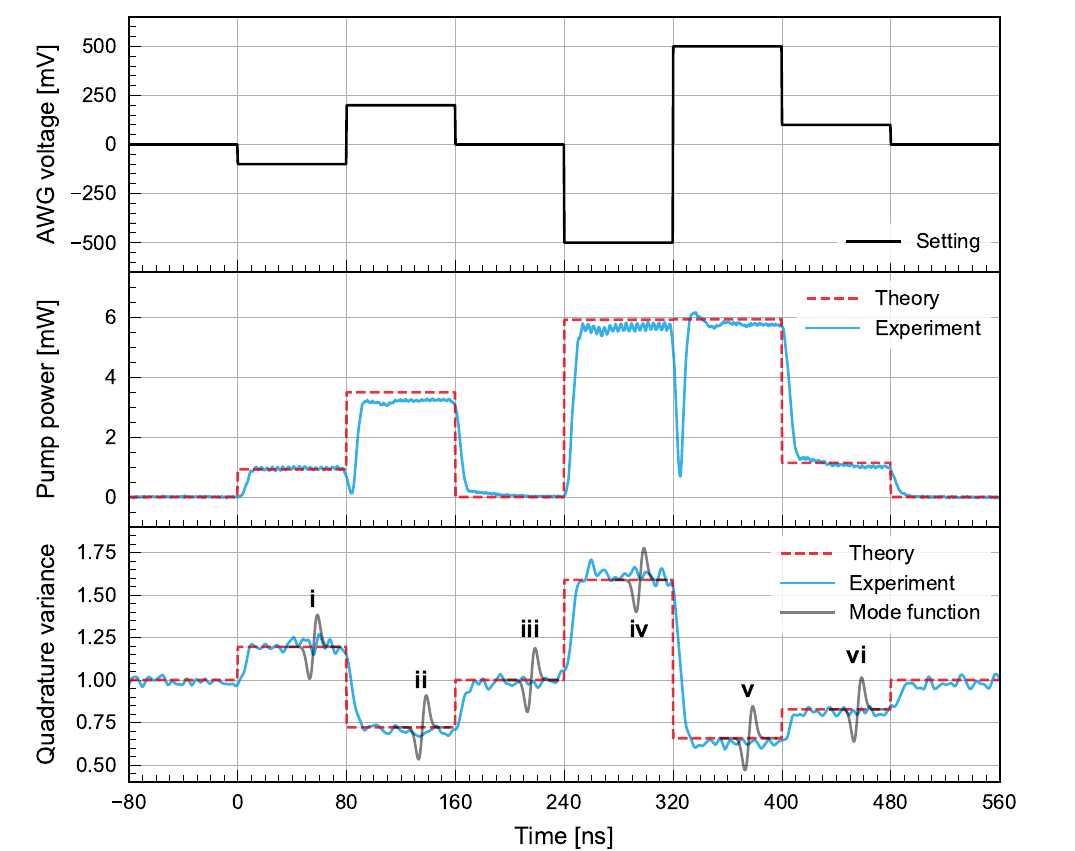}
	  \subcaption{}\label{pic:3a}
	  \vspace{3pt}
	\end{minipage}
	\begin{minipage}[t]{\linewidth}
	  \centering
	  \includegraphics{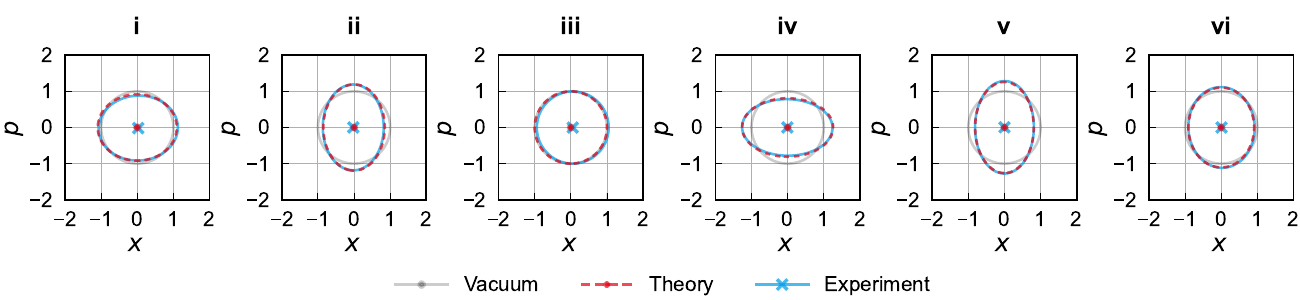}
	  \subcaption{}\label{pic:3b}
	\end{minipage}
	\caption{Generation of time-multiplexed squeezed states.
	(a) It was obtained in the same way as Fig.~\ref{pic:2}.
	The mode functions of the evaluated wave-packet modes are also shown.
	\add{Here it is noted that we define these functions within the homodyne detector bandwidth.}
	(b) Each panel shows the Wigner function of each wave-packet mode where each Roman numeral corresponds to that in the lower panel of (a).
	Here the quadrature amplitude $\hat{x}$ is the quadrature whose variance is plotted in the lower panel of (a) and we set $\hbar=2$.
	The o- and x-shaped symbols are the centers of the Wigner function, while the ellipses are contour lines where the value of the Wigner function is $1/\sqrt{e}$ relative to that of the center.
	The standard errors of $\ev{\hat{x}}, \ev{\hat{p}}, \Delta x$, and $ \Delta p$ are typically less than 0.01.
	}\label{pic:3}
\end{figure}

Next, to demonstrate that our light source can be applied to the TM QIP experiments, we confirmed that it can programmably produce time-multiplexed squeezed states.
The time-multiplexed squeezed states represent sequential squeezed pulses arranged in the time domain, where each pulse has different squeezing levels and phases in general.
The typical pulse intervals were several tens to hundreds of nanoseconds in the previous CV TM experiments~\cite{yokoyama_ultra-large-scale_2013, yoshikawa_invited_2016,larsen_deterministic_2019, asavanant_generation_2019, larsen_fiber-coupled_2019, larsen_deterministic_2021,asavanant_time-domain-multiplexed_2021,takeda_-demand_2019,enomoto_programmable_2021,madsen_quantum_2022}.
In this section, we generate the time-multiplexed squeezed states in the above time scale by setting staircase AWG voltage and evaluate them with quantum tomography.

To verify the operation of our light source, we set the pulse intervals to \SI{80}{\nano\second} and varied the AWG voltage every \SI{80}{\nano\second} like the upper panel in Fig.~\ref{pic:3a} as one example.
The reasons why we chose \SI{80}{\nano\second} are as follows.
First, the time width of the wave-packet mode was set to \SI{30}{\nano\second} so that the mode was included within the homodyne detector bandwidth, $\sim$\SI{200}{\mega\hertz}.
Second, as shown in Fig.~\ref{pic:2a}, the rise and fall time of the pump power was almost \SI{10}{\nano\second}, respectively.
In addition, we found that the pump power shows the ringing behavior after the rising edge for several tens of nanoseconds when the AWG voltage is large (see the middle panel of Fig.~\ref{pic:3a} and the second panel from the top of Fig.~\ref{pic:4}).
Hence, we prepared a time margin of \SI{50}{\nano\second} between neighboring wave-packet modes.
For these reasons, we set the period of the staircase AWG voltage to \SI{80}{\nano\second}, the summation of \SI{50}{\nano\second} and \SI{30}{\nano\second}.
If we neglect the small ringing effect, we can set the period shorter.

The middle and lower panels in Fig.~\ref{pic:3a} show the time variation of the pump power and quadrature variance like Fig.~\ref{pic:2}, respectively.
One difference from Fig.~\ref{pic:2} is that, to vary the pump power and squeezing level over a wide range, we used a large AWG voltage beyond the range where the linear relationship between the pump power and the AWG output power is valid.
Outside the linear range, the correspondence between the pump power and AWG output power had been measured in advance.
The experimental and theoretical curves agree well.

To characterize the output squeezed states in more detail, we acquired the quadratures at multiple phases and evaluated the generated states by quantum tomography.
We changed the LO phase with \SI{15}{\degree} intervals at 12 phases and got \num{5000} frames of time-series quadratures at each phase.
To evaluate the wave-packet modes, we here defined the mode functions as follows:
\begin{align}
	f(t) = 
	\begin{dcases}
		t e^{-\gamma^2 t^2}  & \text{if} \quad |t|\leq \tfrac{t_w}{2}\\
		0 & \text{otherwise}
	\end{dcases},
\end{align}
where $\gamma=\SI{2.5e8}{\per\second}$, $t_w=\SI{30}{\nano\second}$, and the normalization is neglected.
Although we can choose an arbitrary temporal mode in this experiment, we used the above function, which was often applied in previous studies due to its robustness to low-frequency noise~\cite{yoshikawa_invited_2016, enomoto_programmable_2021}.
We calculated $5000\times12$ quadratures of each wave-packet mode in $f(t-t_c)$.
Here $t_c$ is a center position of each temporal mode and modes were set at \num{80}-\si{\nano\second} intervals.
The lower panel of Fig.~\ref{pic:3a} shows the shape and horizontal position of the mode functions.
We split \num{5000} samples into \num{10} sets to estimate the standard errors.
We performed quantum tomography for each set by the maximum likelihood estimation method, assuming that the generated states were Gaussian states~\cite{mauro_dariano_parameter_2000}.
These results are plotted as the Wigner functions in Fig.~\ref{pic:3b}.
We defined the quadrature amplitude at the default phase as $\hat{x}$ and its orthogonal quadrature amplitude as $\hat{p}$, where we set $\hbar=2$.
Based on the theoretical quadrature variances obtained in the above-described way, we also drew the theoretical ellipses of the Wigner functions with the center at the origin and the experimental ellipses agree well with the theoretical ones.
In each wave-packet mode, the angle deviation between the experimental and theoretical ellipse ranges from \SIrange{-1}{3}{\degree} except the vacuum (Fig.~\ref{pic:3b} iii), which suggests that we can control the squeezing phase accurately as expected.
The results prove that we can switch the squeezing levels and phases of wave-packet modes every \SI{80}{\nano\second}.

It is noted that Fig.~\ref{pic:3a} is just an example and we can vary the output interval and pattern of the squeezed pulses by only changing the AWG output on the computer.
We have already confirmed that our system runs properly also in the case of other time intervals and patterns.
Here the time intervals are comparable to the typical wave-packet interval of the previous TM experiments; hence our programmable light source can be applied in a wide variety of TM QIP experiments.

\subsection{Generation of EPR state in frequency domain}\label{sec:Generation of frequency-entangled states}
Even though our main motivation is to realize a programmable time-multiplexed squeezed light source for QIP in the time domain, this source also has the potential to produce useful quantum states in the frequency domain.
Here, we demonstrate this potential by generating an Einstein-Podolsky-Rosen (EPR) state~\cite{ou_realization_1992} in the frequency domain as the simplest example.
It is achieved by modulating the pump light appropriately and choosing the corresponding proper temporal modes.

\begin{figure}[htp]
	\centering
	\includegraphics[scale=0.5]{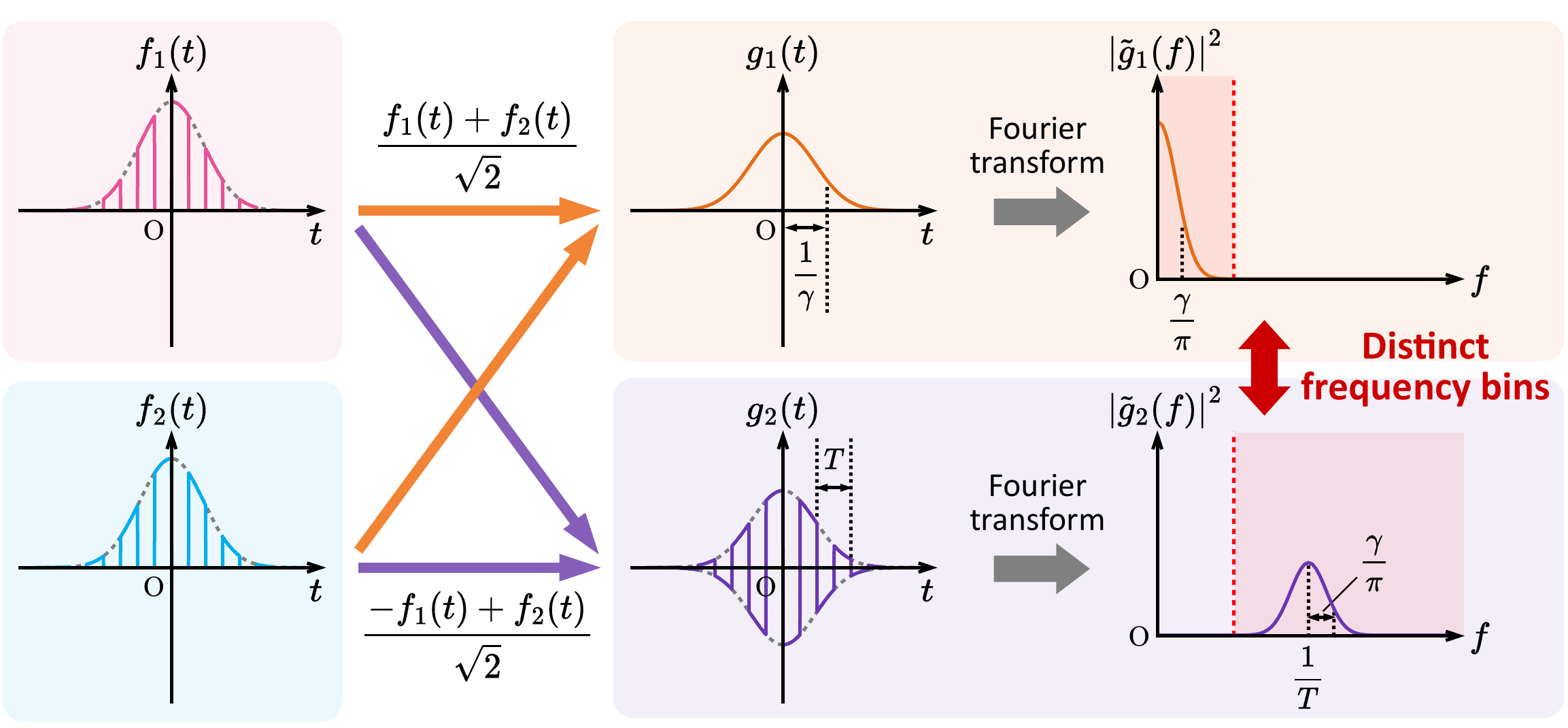}
	\caption{Generation method of EPR state in frequency domain.
	By modulating the pump light appropriately and considering corresponding mode functions $g_1(t)$ and $g_2(t)$, we can produce the EPR state in two distinct frequency bins.
	Fourier transform functions are denoted by a tilde.
	Though $|\tilde{g}_2(f)|^2$ have more peaks of higher frequency out of the plot range, the frequency components around $1/T$ plotted in the figure are dominant.}
	\label{pic:4-0}
\end{figure}

We first explain how to make the EPR state in the frequency domain.
We define two orthogonal wave-packet modes as follows:
\begin{align}
	f_1(t) &= 
	\begin{dcases}
		C e^{-\gamma^2 t^2} h(t;1,0) & \text{if} \quad |t|\leq \tfrac{t_w}{2}\\
		0 & \text{otherwise}
	\end{dcases},
	\label{eq:f1}
	\\
	f_2(t) &= 
	\begin{dcases}
		C e^{-\gamma^2 t^2} h(t;0,1) & \text{if} \quad |t|\leq \tfrac{t_w}{2}\\
		0 & \text{otherwise}
	\end{dcases},
	\label{eq:f2}
\end{align}
where $C$ is a normalization constant and the square-wave function $h(t;a,b)$ is defined with an integer $n$ as
\begin{align}
	h(t;a,b) &=
	\begin{dcases}
		a & \text{if} \quad nT \leq t <\left(n+\tfrac{1}{2}\right)T\\
		b & \text{if} \quad \left(n+\tfrac{1}{2}\right)T \leq t <(n+1)T\\
	\end{dcases}
	.
	\label{eq:h}
\end{align}
In Eqs.~\eqref{eq:f1} and \eqref{eq:f2}, the following relationships are valid for $|t|\leq t_w/2$: $f_1(t)=0$ when $f_2(t)\neq 0$, and $f_2(t)=0$ when $f_1(t)\neq 0$, as shown in Fig.~\ref{pic:4-0}.
Now suppose that our light source produces $x$-squeezed light and $p$-squeezed light alternately every $T/2$.
In this case, $x$-squeezed light is generated when $f_1(t)\neq 0$ and $p$-squeezed light is generated when $f_2(t)\neq 0$; hence the quantum state in mode $f_1(t)$ is an $x$-squeezed state and that in mode $f_2(t)$ is a $p$-squeezed state.
Here we assume that their squeezing bandwidth is ideally infinite and that the parameters in $f_1(t)$ and $f_2(t)$ meet $\gamma/\pi \ll 1/T$.

Next, we define two new orthogonal modes as
\begin{align}
	g_1(t) &= 
	\begin{dcases}
		\tfrac{C}{\sqrt{2}} e^{-\gamma^2 t^2} & \text{if} \quad |t|\leq \tfrac{t_w}{2}\\
		0 & \text{otherwise}
	\end{dcases},
	\\
	g_2(t) &= 
	\begin{dcases}
		\tfrac{C}{\sqrt{2}} e^{-\gamma^2 t^2} h(t;-1,1) & \text{if} \quad |t|\leq \tfrac{t_w}{2}\\
		0 & \text{otherwise}
	\end{dcases}.
	\label{mode_epr}
\end{align}
These modes have the following relationships: $g_1(t)=\left(f_1(t)+f_2(t)\right)/\sqrt{2}$ and $g_2(t)=\left(-f_1(t)+f_2(t)\right)/\sqrt{2}$, so that modes $g_1(t)$ and $g_2(t)$ correspond to the modes generated by mixing $f_1(t)$ and $f_2(t)$ with a 50:50 beam splitter.
In general, when we input two orthogonally squeezed states into the 50:50 beam splitter, the outputs are the EPR state.
In a similar way, quantum states in modes $g_1(t)$ and $g_2(t)$ are the EPR state.
As Fig.~\ref{pic:4-0} shows, $g_1(t)$ is a Gaussian curve, whereas $g_2(t)$ has the same envelope as $g_1(t)$ but is oscillating in a rectangular shape with the period $T$ within the envelope.
In the frequency region, $g_1(t)$ has a bandwidth related to the parameter $\gamma$ around the frequency \SI{0}{\hertz}, while $g_2(t)$ has almost the same bandwidth around the different frequency determined by the oscillating frequency, $1/T$.
Now under the assumption $\gamma/\pi \ll 1/T$, two modes $g_1(t)$ and $g_2(t)$ are almost divided in the frequency domain.
Thus, by periodically switching $x$- and $p$-squeezed light and considering appropriate modes $g_1(t)$ and $g_2(t)$, we can generate the EPR state in two distinct frequency bins.

\begin{figure}[bp]
	\centering
	\includegraphics{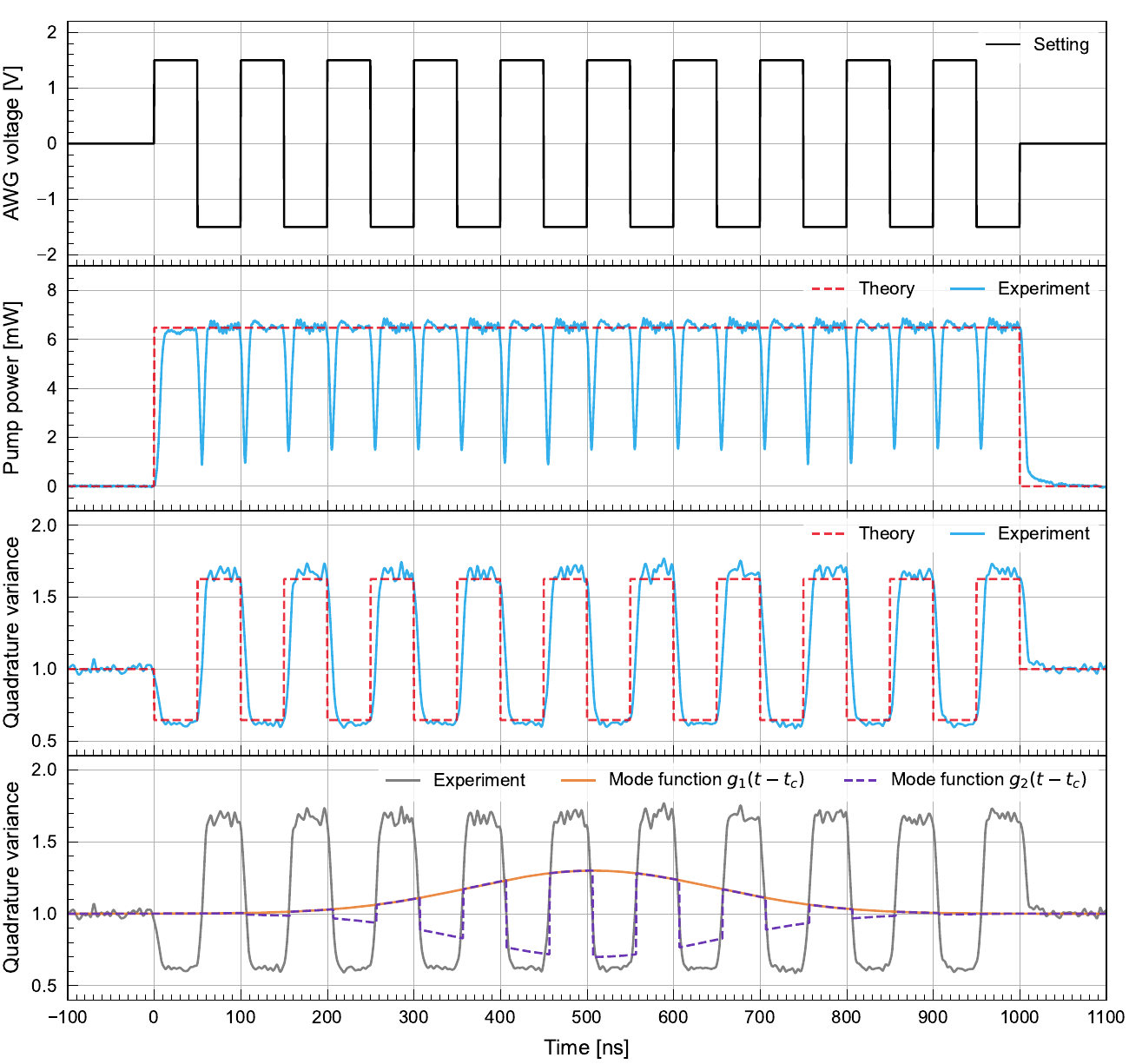}
	\caption{Generation of EPR state in frequency domain.
	Upper three panels were obtained in the same way as Fig.~\ref{pic:3a}.
	% \erase{The lowest panel shows the mode functions.}
	\add{The lowest panel shows the mode functions, which are within our homodyne detector bandwidth in the frequency domain.}
	}
	\label{pic:4}
\end{figure}

Based on the above theory, we produced such an EPR state with our system.
We alternated the positive and negative AWG voltages to generate $x$- and $p$-squeezed light every \SI{50}{\nano\second} ($T=\SI{100}{\nano\second}$) for \SI{1000}{\nano\second} ($t_w=\SI{1000}{\nano\second}$) like the top panel of Fig.~\ref{pic:4}.
\add{Here we determined these experimental parameters so that two modes $g_1(t)$ and $g_2(t)$ are within our homodyne detector bandwidth in the frequency domain.}
Then we acquired \num{5000} frames of time-series quadratures $\hat{x}$ and $\hat{p}$.
The second and third panels from the top show the time variation of the pump power and $x$-quadrature variance in the same way as Fig.~\ref{pic:3a}.
We find that the experimental lines agree well with the theoretical lines.

Next, we evaluated the entanglement of the generated EPR state.
We integrated the not-filtered time-series quadratures $\hat{x}$ and $\hat{p}$ by weighting mode functions $g_1(t-t_c)$ and $g_2(t-t_c)$, where $\gamma=\SI{5e6}{\per\second}$, $T=\SI{100}{\nano\second}$, $t_w=\SI{1000}{\nano\second}$, and $t_c$ is the optimized center position.
These mode functions are shown in the lowest panel of Fig.~\ref{pic:4}.
We denoted the quadrature for $g_1(t-t_c)$ and $g_2(t-t_c)$ as $(\hat{x}_1, \hat{p}_1)$ and $(\hat{x}_2, \hat{p}_2)$, respectively.
From the acquired \num{5000} frames, we obtained \num{5000} samples of each quadrature.
To confirm the entanglement between the two modes, we evaluated the inseparability criterion~\cite{duan_inseparability_2000} and obtained
\begin{align}
	\langle \left[ \Delta (\hat{x}_1 - \hat{x}_2)\right]^2 \rangle
	+ \langle \left[ \Delta (\hat{p}_1 + \hat{p}_2)\right]^2 \rangle
	=2.79 \pm 0.03 < 4\quad (\hbar=2).
	\label{inseparability}
\end{align}
Here the standard error was estimated by dividing \num{5000} quadratures into \num{10} sets and analyzing them separately.
This result meets the inseparability condition that the left side of Eq.~\eqref{inseparability} is under four, thus demonstrating that modes $g_1(t-t_c)$ and $g_2(t-t_c)$ are entangled.
Incidentally, the above value in Eq.~\eqref{inseparability} corresponds to the effective squeezing level of \SI{1.56}{\deci\bel}.
%\erase{It can be compared with the squeezing level of >2dB in Fig.~\protect{\ref{pic:1-2-1}} because the pump power in Fig.~\protect{\ref{pic:4}}, which was monitored after the OPA, almost corresponds to the input pump power of 15mW when Fig.~\protect{\ref{pic:1-2-1}} was obtained.}
\add{It can be compared with the squeezing level of $\SI{>2}{\deci\bel}$ in Fig.~\ref{pic:1-2-1} because the output pump power in Fig.~\ref{pic:4} almost corresponds to that of \SI{6.5}{\milli\watt} when Fig.~\ref{pic:1-2-1} was obtained.}
The effective squeezing level of \SI{1.56}{\deci\bel} is lower than \SI{2}{\deci\bel} because the two modes include the switching regions between $x$- and $p$-squeezed light.

Finally, we mention the frequency components of these two modes.
Their parameters are set to $\gamma=\SI{5e6}{\per\second}$ and $T=\SI{100}{\nano\second}$.
In the frequency domain, $g_1(t-t_c)$ follows the Gaussian distribution around \SI{0}{\hertz} with their half width at half maximum (HWHM) of \SI{1.3}{\mega\hertz}, whereas $g_2(t-t_c)$ also roughly follows the Gaussian distribution around \SI{10}{\mega\hertz} with the same HWHM.
Hence, these two distributions are almost perfectly divided in the frequency domain.
If necessary, these two modes can be spatially split by a frequency filter of \SI{5}{\mega\hertz} cutoff with a separation error of less than 0.0001\%, in principle.

In this way, we experimentally confirmed that our light source can also generate the EPR state in two distinct frequency bins.
This demonstration is just one simple example.
More sophisticated pump modulation may be able to produce more varieties of useful quantum states in both the time and frequency domain, which is left to future work.

\section{Conclusion \add{\& Discussion}}\label{sec:Coclusion}
We developed a programmable time-multiplexed squeezed light source with the original idea of modulating the continuous pump light of the waveguide OPA.
Showing the example of varying the pump power and phase at \num{80}-\si{\nano\second} intervals, we demonstrated that our light source can change each squeezing level and phase of wave-packet modes expectedly at the time interval below \SI{100}{\nano\second}, which is comparable to the time scale of the previous CV TM experiments~\cite{yokoyama_ultra-large-scale_2013, yoshikawa_invited_2016,larsen_deterministic_2019, asavanant_generation_2019, larsen_fiber-coupled_2019, larsen_deterministic_2021,asavanant_time-domain-multiplexed_2021,takeda_-demand_2019,enomoto_programmable_2021,madsen_quantum_2022}.
Besides, in our system, the output timing and pattern can be changed just by a computer program.
Hence, our programmable light source can be applied to various CV QIP tasks based on TM.
Furthermore, by devising the pump modulation, we generated the EPR state in the frequency domain.
This result proves that our system is also useful for CV QIP using the frequency domain.

%\erase{Our light source can be improved more in future work.}
%\erase{Now we use the AOMs for the pump light modulation, which limits the switching speed and the output squeezing level.}
%\erase{These problems will be solved if we introduce an EOM with a faster modulation speed and amplify the pump light.}
%\erase{Besides, although this light source can only produce Gaussian states like the vacuum state and squeezed states, if combined with the photon detectors, non-Gaussian states~\protect{\cite{kawasaki_generation_2022,takase_generation_2022}} can also be generated programmably.}
\add{
We here specifically mention the applications of our light source.
The light source demonstrated in this work has the limited controllability of only low squeezing (up to \SI{\sim2}{\deci\bel}).
In general, QIP tasks based on high-fidelity quantum gates and highly entangled states always require high squeezing beyond the current controllability of our light source.
In contrast, quantum computations and quantum state generation based on GBS~\cite{huh_boson_2015,sempere-llagostera_experimentally_2022,sabapathy_production_2019,fukui_efficient_2022} require tuning the levels of individual squeezed states according to each purpose, rather than always producing highly squeezed light.
The programmable light source in this work is especially suitable for such purposes.
In particular, several experiments have proven that the controllability of low squeezing below \SI{\sim2}{\deci\bel} was already sufficient to perform certain GBS-based computations such as calculating the vibronic spectra of molecules~\cite{huh_boson_2015} and characterizing features of graphs~\cite{sempere-llagostera_experimentally_2022}.
In addition, Ref.~\cite{sabapathy_production_2019} revealed that some useful non-Gaussian states can be generated based on GBS with low squeezing (\SI{\sim3}{\deci\bel}).
The required squeezing level in this GBS-based state generation depends on the target states and circuit configuration, so even lower squeezing may be sufficient for generating other non-Gaussian states.
Therefore, our light source is applicable to implementing such GBS-based tasks in scalable TM schemes.
}

\add{
In future work, our light source can be improved more on the following three points; the output state variety, the output switching speed, and the maximum squeezing level.
First, we can extend the available classes of the output states by introducing photon detectors.
The current system can only produce Gaussian states like the vacuum and squeezed states, but non-Gaussian states~\cite{kawasaki_generation_2022,takase_generation_2022} can also be generated programmably if combined with the photon detectors.
Second, we can improve the switching speed by introducing an EOM with a faster modulation speed for pump modulation system.
In the current setup, the rise time of the AOMs is \SI{6}{\nano\second}, thus limiting the switching speed of the squeezing level.
In contrast, EOMs can easily reach a rise time below \SI{1}{\nano\second}.
Third, we can improve the squeezing level by amplifying the pump power after the AOMs.
The AOMs limit the maximum pump power of the OPA as mentioned in Sec.~\ref{sec:Squeezing spectrum of OPA} and thus limit the maximum squeezing level to \SI{\sim 2}{\deci\bel}.
In the previous study, the same-type waveguide OPA module was demonstrated to produce up to \SI{\sim 6}{\deci\bel} squeezing at the CW pump power of \SI{\sim 500}{\milli\watt} without any degradation due to photorefractive damage~\cite{kashiwazaki_fabrication_2021}.
Furthermore, a similar waveguide showed even higher durability over the CW pump power of \SI{1}{\watt}~\cite{kashiwazaki_over-30-db_2019,kazama_over-30-db_2021} (such power corresponds to the power density of \SI{\sim 1}{\mega \watt \per \centi \metre \squared}, while the durability of the material itself is \SI{120}{\mega \watt \per \centi \metre \squared}~\cite{volk_optical-damage-resistant_1990}).
Thus, our light source is expected to produce \SI{6}{\deci\bel} or even higher squeezing by sufficiently amplifying the pump power after the AOMs with an optical amplifier.
In general, dynamical switching to such high pump power can cause a temperature increase in the waveguide.
The temperature increase will change the phase matching conditions and fiber-coupling efficiency, possibly causing a hysteresis of the squeezing level.
However, in our experiment, the temperature increase of the waveguide would be small as long as the number of highly squeezed pulses is not so large.
This is because our system is designed to operate at low constant pump power for most of the operating time and switch to high pump power only for a short period (\SI{\sim 100}{\nano\second}, which corresponds to a 0.02\% duty cycle) when highly squeezed pulses are necessary.
Even if we sequentially inject many high-power pump pulses and such a hysteresis is caused, our light source can cancel out its effect by adopting a control sequence that takes the hysteresis into account.
}

Our light source will be a key device to enable programmable QIP tasks such as one-way quantum computation~\cite{yoshikawa_invited_2016,larsen_deterministic_2019,asavanant_generation_2019,larsen_deterministic_2021,asavanant_time-domain-multiplexed_2021}, loop-based quantum computation~\cite{takeda_universal_2017,takeda_-demand_2019,enomoto_programmable_2021}, GBS~\cite{madsen_quantum_2022}, quantum reservoir computing~\cite{garcia-beni_scalable_2022}, and so on.
Moreover, due to its versatility, our light source can also be employed in various experiments of quantum optics including quantum sensing~\cite{lawrie_quantum_2019} and quantum communication~\cite{flamini_photonic_2019}.

\begin{backmatter}
\bmsection{Funding}
Japan Science and Technology Agency (JPMJMS2064); Japan Society for the Promotion of Science (20H01833, 21K18593).

\bmsection{Acknowledgments}
S.~T. acknowledges supports from MEXT Leading Initiative for Excellent Young Researchers, Toray Science Foundation (19-6006), Matsuo Foundation, and the Canon Foundation. H.~T. acknowledges a financial support from The Forefront Physics and Mathematics Program to Drive Transformation (FoPM). The authors thank Takahiro Mitani for the careful proofreading of the manuscript.

\bmsection{Disclosures}
The authors declare no conflicts of interest.

\bmsection{Data Availability Statement}
Data underlying the results presented in this paper are not publicly available at this time but may
be obtained from the authors upon reasonable request.

% \bmsection{Supplemental document}

\end{backmatter}

%%%%%%%%%%%%%%%%%%%%%%% References %%%%%%%%%%%%%%%%%%%%%%%%%

%%%%%%%%%% If using BibTeX:
\bibliography{PTMSLS_arXiv_final}

%%%%%%%%%% If preparing manually:
% \newcommand{\enquote}[1]{``#1''}

% \bibitem{Zhang:14}
% Y.~Zhang, S.~Qiao, L.~Sun, Q.~W. Shi, W.~Huang, L.~Li, and Z.~Yang,
%   \enquote{Photoinduced active terahertz metamaterials with nanostructured
%   vanadium dioxide film deposited by sol-gel method,}
%   {\protect\JournalTitle{Optics Express}} \textbf{22}, 11070--11078 (2014).

% {Optical Society}, \enquote{{OSA Publishing},}
%   \url{http://www.osapublishing.org}.

% \bibitem{FORSTER2007}
% P.~Forster, V.~Ramaswamy, P.~Artaxo, T.~Bernsten, R.~Betts, D.~Fahey,
%   J.~Haywood, J.~Lean, D.~Lowe, G.~Myhre, J.~Nganga, R.~Prinn, G.~Raga,
%   M.~Schulz, and R.~V. Dorland, \enquote{Changes in atmospheric consituents and
%   in radiative forcing,} in \enquote{Climate Change 2007: The Physical Science
%   Basis. Contribution of Working Group 1 to the Fourth assesment report of
%   Intergovernmental Panel on Climate Change,}  S.~Solomon, D.~Qin, M.~Manning,
%   Z.~Chen, M.~Marquis, K.~B. Averyt, M.~Tignor, and H.~L. Miler, eds.
%   (Cambridge University Press, 2007).

% \end{thebibliography}

\end{document}